# Temperature dependence of micelle shape transitions in copolymer solutions: the role of inter-block incompatibility

M. J. Greenall[a] and M. J. Derry[b]

[a]School of Mathematics and Physics, University of Lincoln, Brayford Pool, Lincoln, LN6 7TS, United Kingdom

[b]Aston Institute for Membrane Excellence, Aston University, Birmingham, B4 7ET, United Kingdom.

The nature of the transition between worm-like and spherical micelles in block copolymer dispersions varies between systems. In some formulations, heating drives a transition from worms to spheres, while in other systems the same transition is induced by cooling. In addition, a sphere-worm interconversion can be accompanied either by an increase or a decrease in the solvation of the core, even if the direction of the temperature dependence is the same. Here, self-consistent field theory is used to provide a potential explanation of this range of behaviour. Specifically, we show that, within this model, the dependence of the transition on the incompatibility $\chi_{BS}$ of the solvophobic block B and the solvent S (the parameter most closely related to the temperature) is strongly influenced by the incompatibility $\chi_{AB}$ between B and the solvophilic block A. When $\chi_{AB}$ is small ($\chi_{AB} \leq 0.1$), it is found that increasing $\chi_{BS}$ produces a transition from worm-like micelles to spheres (or, more generally, from less curved to more curved structures). When $\chi_{AB}$ is above 0.1, increasing $\chi_{BS}$ drives the system from spheres to worm-like micelles. Whether a transition is observed within a realistic range of $\chi_{BS}$ is also found to depend on the fraction of solvophilic material in the copolymer. The relevance of our calculations to experimental results is discussed, and we suggest that the direction of the temperature dependence may be controlled not only by the solution behaviour of the solvophobic block (upper critical solution temperature versus lower critical solution temperature) but also by $\chi_{AB}$.

## Introduction

The morphologies of the self-assembled structures, such as micelles and vesicles, formed by amphiphilic block copolymers in solution can often be controlled by changing the temperature[1]. These transitions have been investigated in a range of systems, with many studies focusing on structures formed by polymerisation-induced self-assembly (PISA)[2–6]. Recent work on thermoresponsive polymers produced by this method[1] demonstrates that the full range of standard structures (spherical micelles, worm-like micelles, vesicles and lamellae) can be accessed in a single system by varying the temperature. Precise control over the self-assembled structures simply on changing the temperature raises the possibility of a range of applications, such as the release of proteins or enzymes encapsulated in a vesicle[1] or the production of an oil that thickens at high temperatures[7].

Studies in this area are complicated by the fact that the nature of the transitions between structures varies significantly from one system to another. In many formulations, heating drives transitions towards more curved structures, for example, from worm-like micelles to spherical micelles[8–13] or from vesicles to worm-like micelles[7, 14].

In other dispersions, the opposite temperature dependence is seen, and heating the system leads to transitions to less curved structures[15, 16], so that sphere-to-worm[17, 18] and worm-to-vesicle[19, 20] transitions are observed as the temperature is increased.

In addition, systems that have the same direction of the temperature dependence can differ in how the core solvation changes during the transition, and recent studies on thermo-responsive block copolymers have found that transitions to less curved structures on heating can be accompanied either by increased or decreased solvation of the core[21, 22].

Self-consistent field theory (SCFT) calculations are a promising route for rationalising the varying temperature dependence observed in micellar systems. Although Monte Carlo methods[23] provide a greater level of microscopic detail, and dissipative particle dynamics techniques[24] give both more detail and information on the dynamics, self-consistent field theory allows a rapid scan of the parameter space[25]. In addition, existing SCFT results[25, 26] show different dependence of the shape transitions on the solvent-solvophobic block chi parameter $\chi_{BS}$, the quantity most usually associated with the temperature[25]. Specifically, depending on the parameters chosen, increasing $\chi_{BS}$ is predicted by SCFT either to drive the system to form flatter structures[25] or to form more curved structures[26]. Our aim here is to find the factors, for example interaction strengths or block lengths, that lead to the difference in the direction of temperature dependence and to interpret our calculations in the light of experimental results. Although much of this experimental data comes from the PISA literature, the model does not depend on the dynamics of this process and should also be applicable to aggregates produced by other methods.

## Method

Self-consistent field theory[27] models polymer molecules as random walks in space. As a mean-field theory, SCFT does not take fluctuations into account and, as an equilibrium model, it finds states that correspond to a local minimum of the free energy. The system is treated as being incompressible, so that the volume fractions of the various species sum to 1 at all points, and the interactions between different molecular species *i* and *j* are assumed to be short ranged with a strength that can be quantified by a single parameter $\chi_{ij}$. These interactions are modelled by position-dependent fields, one associated with each species.

To model micelle formation, a solution of AB diblock copolymers in solvent S is studied in the canonical ensemble, so that the overall volume fractions of all species remain fixed. The degree of polymerisation $N$ of the copolymers, the overall volume fraction $\phi$ of copolymer, the block volume fraction $f_A$ of the copolymer that consists of A monomers and the interaction parameters $\chi_{AB}$, $\chi_{AS}$ and $\chi_{BS}$ are all specified at the beginning of the calculation, and the outputs are the position-dependent volume fractions $\phi_A(\mathbf{r})$, $\phi_B(\mathbf{r})$ and $\phi_S(\mathbf{r})$ and the free energy $F$. Since we are investigating broad trends rather than attempting to model a specific system, the volume of an A segment, the volume of a B segment and the volume of a solvent molecule are all set to the same value, which we write in terms of the segment density $\rho_0$ as $1/\rho_0$.

Details of SCFT applied to copolymers in a solvent can be found in, for example, Ref. 28. Specialising to the current case of AB diblock copolymers, the free-energy density measured with respect to that of a homogeneously mixed system can be written as[28, 29]

$$\frac{FN}{k_B T \rho_0 V} = -\frac{1}{V}\int d\mathbf{r}\, [\chi_{AB} N(\phi_A(\mathbf{r}) - \phi f_A)(\phi_B(\mathbf{r}) - \phi(1-f_A)) \\ + \chi_{AS} N(\phi_A(\mathbf{r}) - \phi f_A)(\phi_S(\mathbf{r}) - (1-\phi)) \\ + \chi_{BS} N(\phi_B(\mathbf{r}) - (1-f_A)\phi)(\phi_S(\mathbf{r}) \\ - (1-\phi))] \\ - \phi \ln(Q_{AB}/V) - (1-\phi)N\ln(Q_S/V). \quad (1)$$

Here, $V$ is the total volume of the calculation box and $k_B T$ is Boltzmann's constant multiplied by the temperature. The calculation of the density profile for a given polymer species involves solving a modified diffusion equation including the field term corresponding to that species, and the solutions of these differential equations are also used to compute $Q_{AB}$, the partition function for a single copolymer chain[30]. The solvent partition function $Q_S$ is calculated directly from the field associated with the solvent[28].

The derivation of the above free energy also gives a set of simultaneous equations that link the fields and densities[28]. Each SCFT calculation is started by making initial guesses for the fields, which are then used to calculate the densities. These densities are then substituted back into the simultaneous equations to update the fields, and the process is repeated until convergence is achieved.

To compute the free energy of a spherical micelle, we work in spherical polar coordinates and assume spherical symmetry of the aggregates, making the problem one-dimensional. Reflecting boundary conditions are imposed at the centre and surface of the system. Once a solution corresponding to a micelle has been computed by the procedure outlined above, the micelle with the lowest free-energy density is found as follows. The volume of the calculation box is adjusted, the SCFT equations are solved again, and the new free-energy density is calculated. This process is continued until a minimum in the free-energy density is located. This is an approximate way of finding the equilibrium state of the whole system, as varying the box size is equivalent to varying the number of micelles[31]. Similarly, worm-like micelles are modelled in cylindrical polar coordinates assuming cylindrical symmetry and no variation in density along the length of the cylinder. This latter assumption implies that the worms are treated as being infinitely long, meaning that, for example, contributions to the free energy from the worm ends or from their curvature are neglected. Finally, Cartesian coordinates are used to investigate bilayers. Again, the problem is made one-dimensional, here by assuming that the density profile only varies perpendicular to the plane of the membrane. As in the cylindrical case, this treats the bilayers as being of infinite extent. This means that edge effects and curvature are not included in the model and no distinction is made between vesicles and lamellae. Given the assumed symmetries, the box volume $V$ in equation 1 must be replaced by an area $A$ in the cylinder case and a length $L$ in the bilayer calculation. The same procedure of varying the box size to find a minimum in the free-energy density is followed in the cylinder and bilayer cases, with the calculation now corresponding to finding an optimum number of cylinders per unit area or bilayers per unit length. To determine which morphology is predicted to be the equilibrium state for a given set of parameters, the free-energy densities are compared.

The diffusion equations are solved using a finite-difference method with a spatial step size of 0.005 in units of the root-mean-square end-to-end distance of the polymers. The dimensionless parameter $s$ that specifies the distance along the polymer backbone varies from 0 to 1 and a step size of 0.0005 is used. The iterative procedure to solve the simultaneous equations involved in SCFT is implemented using simple mixing followed by damped Anderson mixing[32–35].

The implementation of SCFT used here is equivalent to that used by Liaw *et al.*[26] to study micelle and bilayer formation in block copolymer/solvent systems, and we have reproduced a representative sample of their results as a check on our method.

| $N_A$ \| $f_A$ \ $\chi_{AB}$ | 0.05 | 0.075 | 0.1 | 0.125 | 0.15 | 0.175 | 0.2 |
|---|---|---|---|---|---|---|---|
| 70 \| 0.28 | B→C | B→C | C | C | C | S→C | S→C |
| 72 \| 0.288 | B→C | B→C | C | C | C | S→C | S→C |
| 74 \| 0.296 | B→C | C | C | C | S→C | S→C | S→C |
| 76 \| 0.304 | B→C | C | C | C | S→C | S→C | S→C |
| 78 \| 0.312 | B→C | C | C | S→C | S→C | S→C | S→C |
| 80 \| 0.32 | B→C | C | C→S→C | S→C | S | S | S |
| 82 \| 0.328 | C→S | C→S | C→S | S | S | S | S |
| 84 \| 0.336 | C→S | C→S | S | S | S | S | S |
| 86 \| 0.344 | C→S | C→S | S | S | S | S | S |
| 88 \| 0.352 | C→S | C→S | S | S | S | S | S |
| 90 \| 0.36 | C→S | C→S | S | S | S | S | S |
| 92 \| 0.368 | C→S | S | S | S | S | S | S |

**Table 1** The morphologies that form in systems with a range of values for the degree of polymerisation of the solvophilic block $N_A$ and the inter-block chi parameter $\chi_{AB}$. The block volume fraction $f_A$ of A-blocks in the copolymer is also given in the first column. The first letter (S, C or B) represents the aggregate shape (sphere, cylinder or bilayer) that forms at lower $\chi_{BS}$, and any subsequent letters give the shapes that form as $\chi_{BS}$ is increased. The behaviour of the four systems marked in grey is discussed further in Figures 1—4. In all cases, the overall polymer volume fraction $\phi = 0.05$ and the total degree of polymerisation $N = 250$.

## Results and discussion

The technique outlined in the preceding section is now applied to a range of systems. Since our focus is on the effect of the properties of the polymer molecules and not their concentration, the polymer volume fraction $\phi$ is kept the same in all calculations. A value of $\phi = 0.05$ is chosen, corresponding to a relatively dilute system. This choice is made because the edge of the calculation box would distort the aggregate at high concentrations. To reduce the number of parameters that are varied at once, the overall degree of polymerisation is also chosen to be the same in all computations. The value used, $N = 250$, is moderate in the context of studies of block copolymer self-assembly in solution[25, 26]. We follow Ref. 16 by setting the solvent-solvophilic block chi parameter to 0.48, which is characteristic of a good solvent[26]. The inter-block chi parameter $\chi_{AB}$ is changed from system to system, and is varied from 0.05, corresponding to moderately immiscible polymers, to 0.2, which gives strong segregation. The volume fraction $f_A$ of each copolymer that consists of solvophilic A monomers is also varied between systems, as this parameter has a strong influence on the shape of the aggregates[36]. The solvent-solvophobic block chi parameter $\chi_{BS}$ is varied *within* each system to mimic the effect of temperature variation[25, 26]. Since both $\chi_{AB}$ and $\chi_{AS}$ are assumed not to vary with temperature, this approach assumes that $\chi_{BS}$ is much more responsive to temperature variations than the first two chi parameters.

Each system shown in Table 1 has a specified value of the inter-block chi parameter $\chi_{AB}$ and the solvophilic block degree of polymerisation $N_A$. Since the total degree of polymerisation is fixed at 250, the degree of polymerisation of the solvophobic block is $N_B = 250 - N_A$ and the solvophilic volume fraction of the copolymers is $f_A = N_A/250$. To focus the study, we concentrate on a range of parameters where spheres and cylinders are the most common morphologies. For each system, the solvent-solvophobic block chi parameter $\chi_{BS}$ is increased from 0.7 to 1.5, so that the core changes from strongly to weakly solvated[26]. As $\chi_{BS}$ is changed, any shape transitions that occur are recorded, with the first letter in each cell in Table 1 representing the morphology (sphere, cylinder or bilayer) formed at low $\chi_{BS}$ and any subsequent letters representing the morphologies that form as $\chi_{BS}$ is increased.

In two regions of Table 1, only one morphology (sphere or cylinder) forms over the range of $\chi_{BS}$ used. However, there are two further regions where a transition between cylinders and spheres is predicted. Specifically, for lower values of $\chi_{AB}$ and higher values of $N_A$ (bottom left of the table), a transition from cylinders to spheres is seen as $\chi_{BS}$ is increased. For higher values of $\chi_{AB}$ and lower values of $N_A$ (top right), the predicted transition runs in the opposite direction, from spheres to cylinders as $\chi_{BS}$ increases. In the top left of Table 1, where $N_A$ and $\chi_{AB}$ are both relatively small, there is a transition from bilayers to cylinders as $\chi_{BS}$ is increased.

These results suggest that the most important quantity in determining the direction of the dependence of the shape transitions on $\chi_{BS}$ could be the inter-block chi parameter $\chi_{AB}$. However, it may be necessary to adjust the block lengths as well as $\chi_{AB}$ for the reversed transitions to be seen. For example, for $N_A \geq 82$, increasing $\chi_{AB}$ moves away from the cylinder-sphere transition to a region of parameter space where only one shape (the sphere) is seen. For the opposite (sphere-cylinder) transition to be observed, it is also necessary to lower $N_A$ to move into the top right of the table.

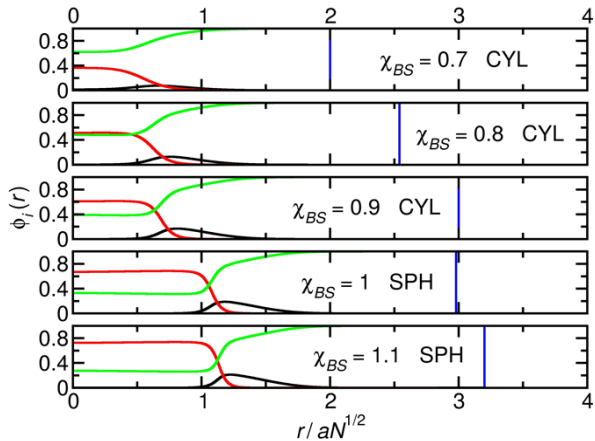

**Fig. 1** Cuts through the density profiles of the micelles that form when $\chi_{AB} = 0.05$ and $N_A = 84$. Black lines: A-blocks (shell); red lines: B-blocks (core); green lines: solvent. Vertical blue lines: boundary of the calculation box.

When $N_A = 80$ and $\chi_{AB} = 0.1$, there is a single system where two transitions occur: from cylinders to spheres and then from spheres back to cylinders. This system is on the border between the regions where cylinder-sphere and sphere-cylinder transitions occur, and the competition between the two morphologies is finely balanced. This re-entrant effect has also been predicted in mean-field calculations that use a single chi parameter[37], and the current calculations give some further information on the conditions for which this phenomenon might occur, although it must be noted that extremely fine tuning of the parameters would be needed.

To gain insight into the processes that lead to the transitions between spheres and cylinders predicted by the model, we plot cuts through the density profiles of the micelles that form as $\chi_{BS}$ is varied for three different systems. These are chosen to lie on a diagonal line starting on the left of the table in the region where cylinder-sphere transitions occur as $\chi_{BS}$ is increased and finishing on the top right where the transitions occur in the opposite (sphere-cylinder) direction. Since $N$ is the same in all systems, distances may be measured in units of the root-mean-square end-to-end distance $aN^{1/2}$ without introducing inconsistency between the figures. In each of Figs 1—3, the local volume fractions of the A, B and S species are plotted. As described in the methods section, the volume of the calculation box is varied to find the structure with the lowest free energy density, and the boundary of this box is marked to show that the aggregates are surrounded by solvent and not distorted by the edge of the system.

In the first system (Fig. 1), $\chi_{AB} = 0.05$ and $N_A = 84$. Here, the relatively large shell (A) block favours more curved structures. As $\chi_{BS}$ increases from 0.7 to 0.9, the radius of the cylinder increases, decreasing its curvature. This compresses the A-blocks in the shell into a progressively more unfavourable state. When $\chi_{BS}$ reaches 1, the system moves from the cylinder to the sphere phase, relaxing the A-blocks. Changing from cylinders to spheres increases the total interfacial area of the micelle cores. However, the system can pay the associated free-energy penalty due to the relatively low value of $\chi_{AB}$. The core blocks also become more compressed when the micelles become spherical, but these are short compared to the other systems studied and the energy cost is again not sufficient to prevent the transition.

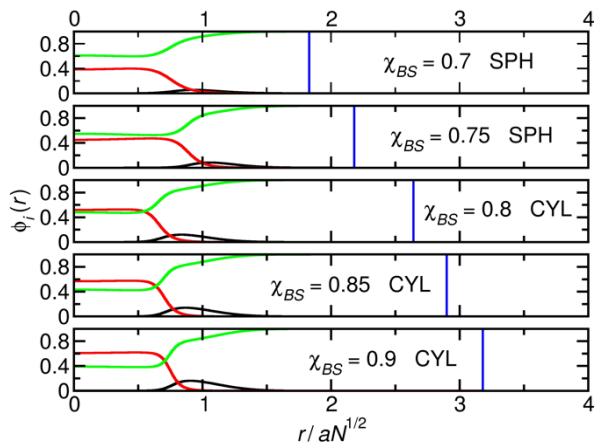

**Fig. 2** Cuts through the density profiles of the micelles that form when $\chi_{AB} = 0.2$ and $N_A = 72$. Black lines: A-blocks (shell); red lines: B-blocks (core); green lines: solvent. Vertical blue lines: boundary of the calculation box.

We now turn our attention to the system (Fig. 2) at the far end of the diagonal cut through the table and contrast it with the previous case. Now, $\chi_{AB} = 0.2$ and $N_A = 72$, and the larger value of $\chi_{AB}$ means that the surface energy is a more important factor in determining the shape of the aggregates. This means that, as $\chi_{BS}$ increases, the system changes from the sphere to the cylinder phase to reduce the overall surface area of its solvophobic cores. This leads to a compression of the shell A-blocks, but these are relatively short, and the associated energy penalty is not enough to keep the system in the sphere phase. The relatively long core blocks also favour the transition to a less curved structure.

The third system we focus on is that where the system changes from cylinders to spheres and then back again (Fig. 3). Here, $\chi_{AB} = 0.1$ and $N_A = 80$, intermediate between the values in the earlier systems, and there is a fine balance between the two phases. It is possible that a factor in the transition back to the cylinder phase at $\chi_{BS} = 1.5$ is the unfavourable state of the B-blocks in the swollen core of the sphere phase as $\chi_{BS}$ increases.

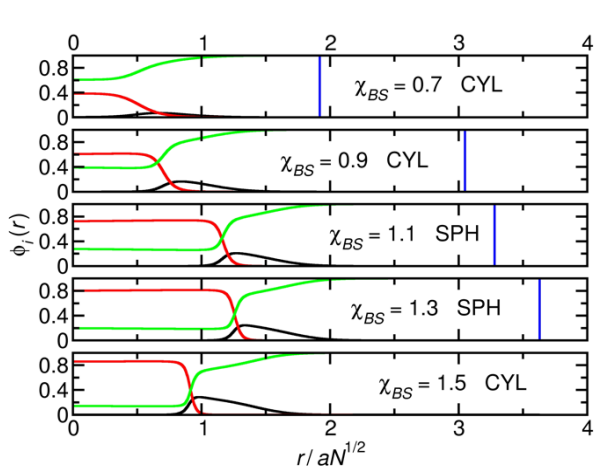 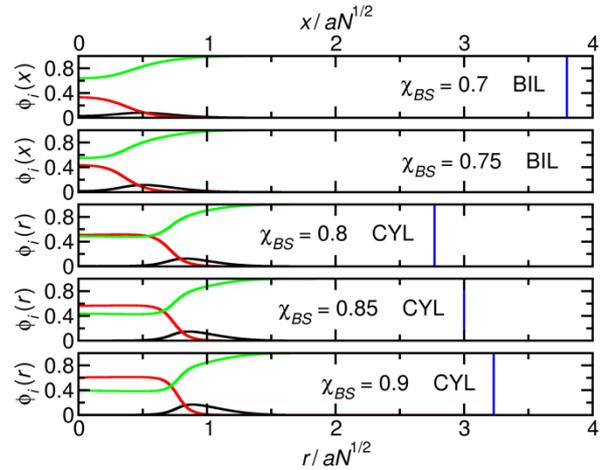

**Fig. 3** Cuts through the density profiles of the micelles that form when $\chi_{AB} = 0.1$ and $N_A = 80$. Black lines: A-blocks (shell); red lines: B-blocks (core), green lines: solvent. Vertical blue lines: boundary of the calculation box. In the final panel, the boundary line lies at $r = 4.59aN^{1/2}$ and is not plotted to make the range of $r$ compatible with Figs 1 and 2.

**Fig. 4** Cuts through the density profiles of the aggregates that form when $\chi_{AB} = 0.05$ and $N_A = 70$. Black lines: A-blocks (shell); red lines: B-blocks (core), green lines: solvent. Vertical blue lines: boundary of the calculation box. In the second panel, this boundary lies at $x = 4.65aN^{1/2}$ and is not plotted.

We now discuss the potential connections between our calculations and experimental results. On the left-hand side of Table 1, where $\chi_{AB} < 0.1$, the system moves from worms to spheres or from bilayers to worms, i.e., from less curved to more curved structures, as $\chi_{BS}$ is increased. If the value of $\chi_{BS}$ decreases on heating, that is, if the core block of the polymer displays upper critical solution temperature-like (UCST-like) behaviour, this corresponds to a transition to less curved structures as the temperature is increased. This is in line with the results of Liaw et al.[26], and we suggest that it may correspond to other systems where the core progressively and uniformly becomes more solvated as flatter structures are formed[1]. This behaviour has been suggested[21, 22] to arise from ingress of the solvent into the centre of the core at higher temperatures, and the current calculations can be interpreted as being in line with this. Specifically, as the system changes from spheres to cylinders as $\chi_{BS}$ is lowered (which we are currently associating with a rise in temperature) in Fig. 1, the interfacial region becomes progressively less sharp, and the solvent penetrates further into the core of the micelle to the extent that, when $\chi_{BS} = 0.7$, the separate and relatively less solvated core region seen in the spheres is no longer present. This effect is seen more clearly in calculations on the transition between cylinders and bilayers that occurs when $\chi_{AB} = 0.05$ and $N_A = 70$ (the top left corner of Table 1). These profiles are shown in Fig. 4, where there is a contrast between the relatively sharply defined core and interface regions in the cylindrical cases and the more diffuse core of the bilayers.

If the core blocks of the polymers display UCST-like behaviour but $\chi_{AB}$ is such that the system lies on the right-hand side of Table 1, then SCFT predicts that the temperature dependence of the shape transitions will be reversed, i.e., that increasing temperature will cause progressively more curved structures to form. This form of temperature dependence has been linked with surface plasticisation, i.e., the penetration of solvent into the surface of the aggregate core[7, 11, 14], which makes the core surface layer behave like an extension of the shell, causing a transition to more curved structures. Again, the density profiles calculated from SCFT can be interpreted as corroborating this. In Fig. 2, where $\chi_{AB} = 0.2$ and $N_A = 72$, the system moves from cylinders to spheres as $\chi_{BS}$ is lowered, which is still being associated with a rise in temperature. As in Fig. 1, the interfacial region becomes broader as $\chi_{BS}$ is reduced, but here a separate core region remains, even at $\chi_{BS} = 0.7$.

In a third class of systems[16], the system transitions to less curved aggregates as the temperature is increased, but the core becomes *less* solvated as this happens. We suggest that these systems lie on the right of Table 1, but that, as noted elsewhere[21, 22], the dependence of $\chi_{BS}$ on temperature is lower critical solution temperature (LCST)-like so that $\chi_{BS}$ increases on heating. This behaviour corresponds to Fig. 2, but with increasing $\chi_{BS}$ now associated with increasing temperature. The association between surface plasticisation and the formation of more curved structures has also been made for these systems[21, 22], and (as noted above) the solvophobic cores of the spheres in Fig. 2 show a broad surface region that remains distinct from the centre of the core.

The above calculations of the density profiles are encouraging, but it should be noted that the solvation of the core and the surface are strongly connected in SCFT, with the broadening of the solvated layer at the surface of the core that we have tentatively associated with surface plasticisation being accompanied by increased solvation of the centre of the core in all results presented here. Furthermore, although the change from the interfacial to the core region is smoothed out in the cases (such as the top panel of Fig. 4) where the solvent penetrates strongly into the centre of the core, the core solvation does not have the uniform profile suggested in, for example, Ref. 20 These considerations mean that the distinction between plasticisation of the surface *versus* uniform plasticisation of the core made in discussions of experimental data is not very marked in the model. In addition, the fact that a separate core region is present in (for example) the cylinders in Fig. 4 but not in the bilayers could simply result from the fact that the bilayer core thickness is less than the core diameter of the cylinders.

## Conclusions

We have shown that self-consistent field theory predicts that the shapes of the aggregates that form in AB block copolymer solutions are strongly dependent on the inter-block incompatibility $\chi_{AB}$ of the copolymers. The importance of the solvophobic/solvophilic block interactions in driving transitions between spheres and worms has been noted in recent experiments[38] and simulations[39], and our results show the utility of SCFT in investigating this effect over a wide range of parameters.

The current work has also demonstrated that the dependence of the shape transitions on the solvophobicity $\chi_{BS}$ of the core-forming B-block can be reversed within SCFT by varying $\chi_{AB}$, with increasing $\chi_{BS}$ driving transitions from less to more curved structures for lower $\chi_{AB}$ and from more to less curved structures for higher $\chi_{AB}$. At an intermediate value of $\chi_{AB}$, a re-entrant transition[37] (cylinders-spheres-cylinders) as a function of $\chi_{BS}$ has been found. From a technical point of view, these results demonstrate the importance of setting $\chi_{AB}$ carefully in SCFT calculations. It is also possible to rationalise a range of experimental observations in terms of our results by also considering the solution behaviour (UCST versus LCST) of the core blocks of the copolymers.

Within SCFT, several extensions of the current work are possible. Firstly, instead of varying the block ratio while keeping the overall degree of polymerisation constant, it could be easier to make a detailed comparison with experiments, especially those involving polymerisation-induced self-assembly[20] if the length of one block (typically, the solvophilic stabiliser) were to be held constant with the other being varied. The parameters in the model would also need to be set more carefully for comparison with a given experiment. As it stands, the three $\chi$ parameters have been treated as independent of each other, and the temperature dependence of only one of these parameters ($\chi_{BS}$) has been considered. Including experimental information on the flexibility of the polymer molecules and allowing this to differ between the two blocks would also be desirable.

Since many recent investigations into micelle and vesicle formation in copolymer solutions involve polymerisation-induced self-assembly, it would also be useful to model the dynamics of this process, which can cause shifts in the phase boundaries[40]. Such more detailed studies could be guided by the parameters identified in the present work and could allow experimentalists to identify formulations with the thermal response required by a specific application in advance.

## Acknowledgements

The authors thank S. J. Hunter and P. D. Topham for useful discussions and comments on the manuscript.

## Notes and references